\documentclass[12pt]{iopart}
\usepackage{graphicx}

\begin{document}

\title{Numerical estimate of infinite invariant densities: application to 
Pesin-type identity}

\author{Nickolay Korabel$^{1}$ and Eli Barkai$^2$}

\address{$^1$ School of Natural Sciences, University of California, Merced, California 95343, USA  \\ $^2$ Physics Department, Institute of Nanotechnology and Advanced Materials, Bar-Ilan University, Ramat-Gan 52900, Israel}
\ead{nnkorabel@gmail.com, barkaie@mail.biu.ac.il}
\begin{abstract}
Weakly chaotic maps with unstable fixed points are investigated in the regime where the invariant density is non-normalizable. We propose that the infinite invariant density $\overline{\rho}(x)$ of these maps can be estimated using $\overline{\rho}(x)=\lim_{t \rightarrow \infty} t^{1-\alpha} \rho(x,t)$, in agreement with earlier work of Thaler. Here $\rho(x,t)$ is the density of particles for smooth initial conditions. This definition uniquely determines the infinite density and is a valuable tool for numerical estimations. We use this density to estimate the sub-exponential separation $\lambda_{\alpha}$ of nearby trajectories. For a particular map introduced by Thaler we use an analytical expression for the infinite invariant density to calculate $\lambda_{\alpha}$ exactly, which perfectly matches simulations without fitting. Misunderstanding which recently appeared in the literature is removed. 


\end{abstract}

\vspace{2pc}
\maketitle

\section{Introduction}\label{Intro}

For a chaotic system with a normalized invariant density Pesin's theorem states the identity between the sum of positive Lyapunov exponents and its Kolmogorov-Sinai (KS) entropy (see \cite{Dor} for conditions). 
In certain weakly chaotic systems, the dynamics still remains quasi random,
 though the entropy and Lyapunov exponents are zero \cite{GW88,AT93}.
Standard chaotic concepts based on exponential separation of nearby trajectories 
are replaced in these systems with new methods which have drawn considerable attention. 
 Previously, a generalization of Pesin identity was suggested for systems whose invariant densities are not absolutely continuous
along expanding directions \cite{GP} and for systems such as logistic map at the edge of chaos based on Tsallis entropy \cite{Tsallis,Robledo} (see also \cite{Gra05} and \cite{Rob06}). Motivated by a question posed by R. Klages, we found a Pesin-type identity in intermittent weakly chaotic maps \cite{KB09,KB10}. Our work shows that Krengel entropy \cite{Kren} must be used instead of KS entropy which is zero for such systems, and that infinite ergodic theory is the mathematical basis for this identity. 

For weakly chaotic maps, in particular maps with marginally unstable fixed points
(see details below), the infinite invariant measure 
\cite{Pian,Thaler83,Thaler00,AB13,Aaronson,Zwei00,AA10,Akimoto08,Akimoto10,Akimoto12,Kessler,KB12} is an essential tool. Even though  this density is not normalizable, still it can be used to 
describe statistical properties of the dynamical system,
 thus replacing the usual normalizable invariant density. However, besides a few exceptions \cite{Thaler00, Lasota} the invariant density  $\overline{\rho}(x)$ is unknown. Here we provide a simple numerical approach for its estimation.
 
Further, the standard scenario of statistical physics is that for chaotic motion,
 the density $\rho(x,t)$ of ergodic systems will tend in to a normalizable
invariant measure $\rho_{{\rm eq}} (x)$ in the long time limit.
For example $\rho_{{\rm eq}}$ is the Boltzmann measure 
for canonical systems. For applicability of the statistical approach the
invariant density must be reached starting from wide classes of initial 
conditions. For systems with an infinite invariant measure do we have similar behavior? Namely will the density approach an infinite measure starting from different types of initial conditions. Here we investigate this issue numerically, and 
show that different initial states  yield the same estimate for the infinite invariant density. This means that we have a simple method to find the infinite
invariant density, at least on a computer. 
The numerical estimation of the infinite invariant density is important
for many applications, in particular for the estimation of the Krengel entropy and hence, according to our generalized Pesin's identity, the averaged separation of trajectories.
Finally, we demonstrate that our numerical infinite invariant density is in excellent agreement with exact analytical infinite invariant density found by Thaler \cite{Thaler00} for a specific map. Using this exact analytical infinite invariant density we corroborated the validity of our generalized Pesin identity without fitting. This leaves no room for speculations and doubts on our results. 

Recently Saa and Venegeroles (SV) proposed a Pesin identity for the same class of dynamics investigated in our work (e.g. Pomeau-Manneville map) for individual trajectories \cite{SV}. We note that the average of this identity over initial conditions is exactly our previously obtained results \cite{KB09,KB10}. We remove misleading statements, for example SV claimed to have ``corrected'' our results. We clarify the notations used, and show that the core of misunderstanding is a trivial constant multiplying the infinite invariant density. This is a simple matter of definition of the infinite density which is not found for normal systems with finite invariant density since there the normalization condition determines uniquely the multiplicative constant in front of the equilibrium density.

\section{Model and Definitions} 

Similar to our work \cite{KB09,KB10} we study one dimensional maps with one or more unstable fixed points. We consider parameter regime where the system has an infinite invariant density \cite{Pian,Thaler83,Thaler00,Aaronson} soon to be defined. 
The discrete time dynamics is governed by $x_{t+1} = M(x_t)$. A prominent example being the Pomeau-Manneville (PM) map \cite{PM80}
\begin{equation}
M(x_t) = x_t + a x_t^{z} \; | \; mod \; 1, \; \; z\ge1,
\label{Maneq}
\end{equation}
with $0<x_t<1$, $a>0$. A second example studied by Thaler \cite{Thaler02} is
\begin{equation}
\label{Man2}
M(x_t) = x_t + \left\{ \begin{array}{ l  l}
(1-c) (x_t/c)^z, & 0< x_t <c \\
- c \left( (1 - x_t)/(1-c) \right)^z, & c < x_t < 1, 
\end{array}
\right.
\end{equation}
where $0<c<1$. Notice that this map is asymmetric with respect to $x=1/2$ for $c \ne 1/2$.
The first map has one unstable fixed point on $x=0$, while the second has two such points on $x=0$ and $x=1$. 

A third map introduced by Thaler \cite{Thaler00} is
\begin{equation}
M(x_t) = x_t \left[ 1 + \left( \frac{x}{1+x} \right)^{z-2} - x^{z-2} \right]^{-1/(z-2)} \; | \; mod \; 1, \; \; z\ge1.
\label{Thalereq}
\end{equation}
This map is similar to the PM map in the sense that it has one unstable fixed point located at $x_t=0$ and for small $x_t$ it has $M(x_t) \simeq x_t + a x_t^z$ behavior. In this case $a=1$. This map is important since Thaler has obtained its exact analytical infinite invariant density.

For all models we are interested in $z>2$ where the usual Lyapunov exponent and KS entropy are zero. When $1<z<2$
they are positive, the invariant measure normalizable and usual Pesin identity holds. In this case the distribution of finite time Lyapunov exponents provides additional information about the behavior of maps \cite{Artuso09,Artuso}.

For a single trajectory the generalized Lyapunov exponent is defined as \cite{KB09,KB10,AA07}
\begin{equation}
\lambda_{\alpha}(x_0) = \frac{1}{ t^{\alpha}} \sum_{k=0}^{t-1} \ln |M'(x_k)|,
\label{eq1}
\end{equation}
where $0<\alpha= 1/(z-1)<1$ for $z>2$ while $\alpha=1$ for $1<z<2$. The generalized Lyapunov exponent (\ref{eq1}) is a random variable
even in the long time limit. Therefore, averaging $\lambda_{\alpha}(x_0)$ 
over initial conditions $\left< ... \right>$, we focused among other things on the averaged generalized Lyapunov exponent 
\begin{equation}
\left< \lambda_{\alpha} \right> = \left< \frac{1}{t^{\alpha}} \sum_{k=0}^{t-1} \ln |M'(x_k)| \right>,
\label{eq2}
\end{equation}
for $t \to \infty$.

\begin{figure}[t!]
  \centering
  \includegraphics[width=.7\textwidth]{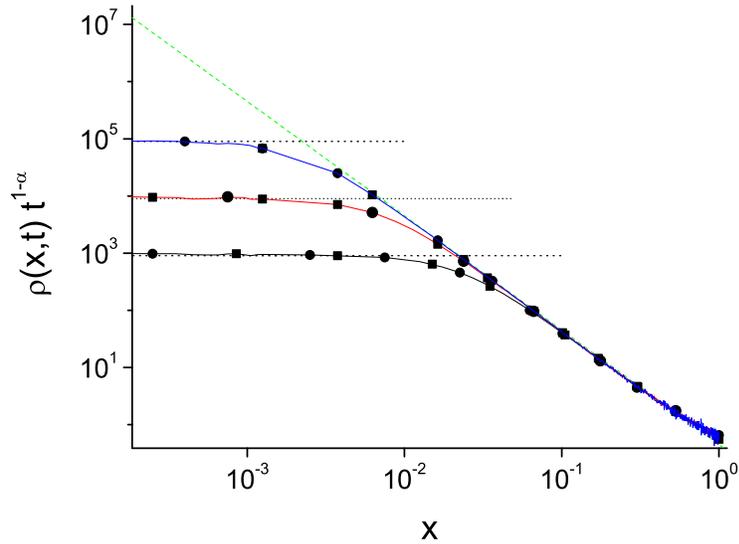}
  \caption{$t^{1-\alpha} \rho(x,t)$ equation (\ref{eq3}) for 
the PM map with $z=3$ ($\alpha=0.5$) and $a=1$. In simulations
the times are $t=10^3, 10^4, 10^5$ (in figure from  
bottom to top). Different initial conditions are used to illustrate that the infinite density is not sensitive to the choice of initial conditions: solid line $\rho(x,0)=1$ for $x \in (0,1)$, circles $\rho(x,0)=2$ for $x \in (0.5,1)$, squares $\rho(x,0)=2 x$ for $x \in (0,1)$. In the limit $t \to \infty$ the system approaches the infinite invariant density: the dashed line  $\overline{\rho}(x) = 0.45 x^{-1/\alpha}$, which is in good agreement with (\ref{rho1}) and (\ref{eqSVb}) without any fitting. As follows from (\ref{g_t}), equation (\ref{rho1}) works for $x \gg x_c$, where $x_c$ is the crossover. For $x\ll x_c$ the finite time $\overline{\rho}(x)$ is correctly described by the second line of equation (\ref{g_t}) (horizontal dotted lines with no fitting). As $t \rightarrow \infty$, $x_c=\alpha^{\alpha} t^{-\alpha} \to 0$ and since $\alpha=1/2$ we have $\overline{\rho}(x) \propto x^{-2}$ when $x\to 0$, which means the system approaches a non-normalizable state.}
  \label{FIG1}
\end{figure}

We defined the infinite invariant density according to \cite{KB09,KB10} (see also Appendix of \cite{AB13} for more discussion of the mathematical aspects of this definition)
\begin{equation}
\overline{\rho}(x) = \lim_{t\rightarrow \infty} \frac{\rho(x,t)}{ t^{\alpha-1}}.
\label{eq3}
\end{equation}
Here $\rho(x,t)$ is the density of particles normalized to unity
$\int_0 ^1 \rho(x,t) {\rm d} x=1$. We later check numerically (see figure \ref{FIG1}) that $\overline{\rho}(x)$ is unique, in the sense that it is independent of the choice of the initial conditions.
 Since $M(x_t)$ conserves normalization, 
$\rho(x,t)$ is normalized for any  $t$ still the integral over 
the limiting value of $\overline{\rho}(x)$ diverges (see \cite{KB09,KB10,KB12}). The conditions and rigorous proof that $\overline{\rho}(x)$ is indeed an invariant density can be found in Thaler's work \cite{Thaler00} for a class of maps with a single unstable fixed point on the origin (see also \cite{AB13}). We adopt this result and use it to find the infinite invariant density numerically.

Using a simple continuous time stochastic model proposed in \cite{Grigo}, we analytically find the approximation for the normalized density of the PM map (\ref{Maneq}) with one unstable fixed point 
\begin{equation}
\label{g_t} 
\rho(x,t) \simeq \left\{ \begin{array}{ l  l}
\frac{a^{\alpha-1} x^{-\frac{1}{\alpha}}}{\alpha^{\alpha}} \; \frac{\sin(\pi \alpha)}{\pi} \; t^{\alpha-1}, & x \gg x_c \\
\frac{\sin(\pi \alpha)}{\pi \alpha}\; \left(\frac{t}{a \alpha}\right)^{\alpha}, & x \ll x_c, 
\end{array}
\right.
\end{equation}
where $\simeq$ denotes small $x$ and long time. The crossover $x_c$ is defined as $\rho_{c \; (x \gg x_c)}(x_c)=\rho_{c \; (x \ll x_c)}(x_c)$. Using (\ref{g_t}) the time dependence of the crossover is obtained $x_c=\alpha^{\alpha} t^{-\alpha}$. Hence, the crossover goes to zero $x_c \rightarrow 0$ as $t \rightarrow \infty$. From definition (\ref{eq3}) we obtain the approximate infinite invariant density for the PM map
\begin{equation}
\overline{\rho}(x) \simeq B x^{-1/\alpha}, \; \; x \rightarrow 0,
\label{rho1}
\end{equation}
and
\begin{equation}
B = \left({ a \over \alpha} \right)^{\alpha-1} {\sin \pi \alpha \over \pi \alpha}. 
\label{eqSVb}
\end{equation}
According to (\ref{rho1}) the slow escape of trajectories from the vicinity of  the unstable fixed point $x=0$
 accumulates the density in its vicinity. The interesting feature being that this
density diverges so strongly on $x\to 0$ that $\overline{\rho}(x)$ is non-normalizable. Importantly Thaler's Theorem \cite{Thaler00} shows that (\ref{eqSVb}) is valid for a large class of maps with a single unstable fixed point on the origin, and which behave like $M(x_t) \sim x_t + a x_t^z$ for $x_t \rightarrow 0$.

Figure \ref{FIG1} demonstrates that when $t \rightarrow \infty$ equations (\ref{rho1}), (\ref{eqSVb}) describe the infinite invariant density for PM map. For finite time $t$ and small $x \ll x_c$ we see deviations in agreement with (\ref{g_t}). Since our theory works for small $x$, not surprisingly (\ref{rho1}) does not work perfectly for $x \simeq 1$, though deviations seem small to the naked eye.
 
\begin{figure}[t!]
  \centering
  \includegraphics[width=.7\textwidth]{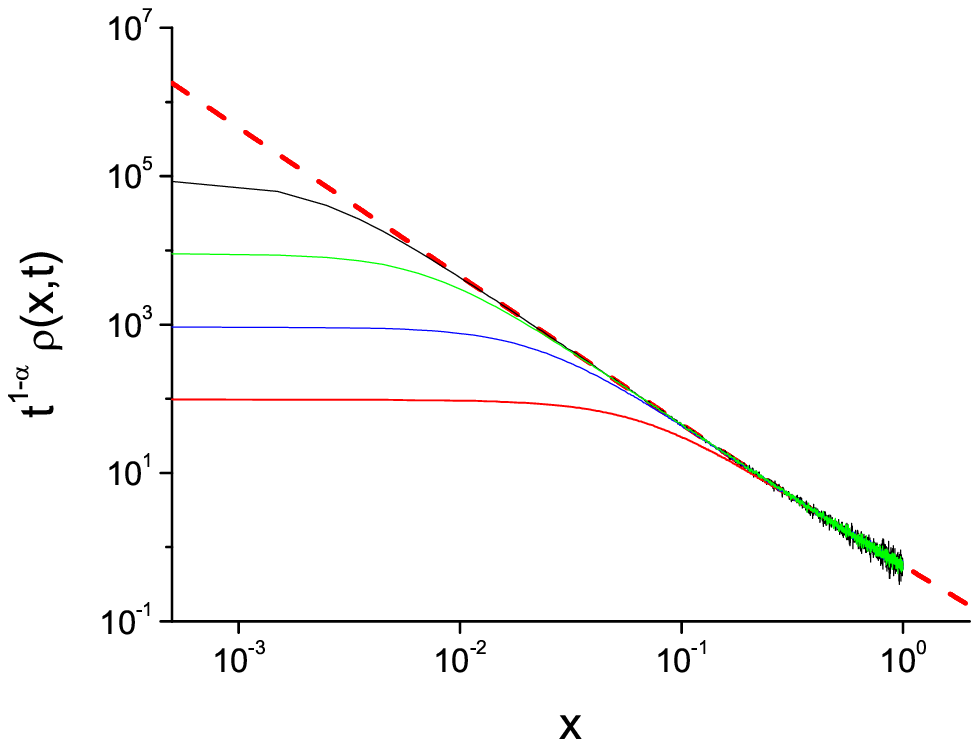}
  \includegraphics[width=.7\textwidth]{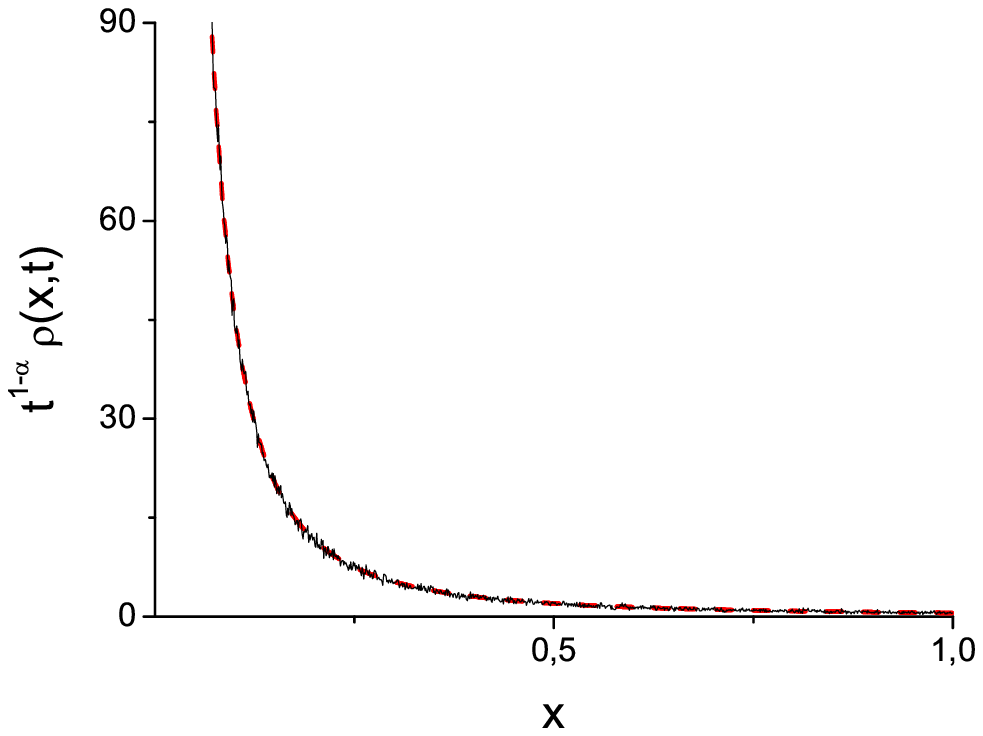}

  \caption{Top:  $t^{1-\alpha} \rho(x,t)$ equation (\ref{eq3}) for 
the Thaler map (\ref{Thalereq}) with $z=3$ ($\alpha=0.5$). In simulations
we have used uniform initial density iterated for $t=10^2, 10^3, 10^4, 10^5$ (in figure from
bottom to top). In the limit $t \to \infty$ the
system approaches the analytical infinite invariant density $\overline{\rho}(x)$ equation (\ref{Thalerinf}) and (\ref{eqSVb}) (the dashed line without any fitting). Bottom: Comparison of $t^{1-\alpha} \rho(x,t)$ for Thaler map (\ref{eq3}) with $z=3$ ($\alpha=0.5$) calculated at $t=10^5$ with analytical infinite invariant density the dashed line $\overline{\rho}(x)$ (\ref{Thalerinf}). Notice the excellent agreement for large range of $x_t$.}
  \label{FIG2NEW1}
\end{figure}

\begin{figure}[t!]
  \centering
  \includegraphics[width=.7\textwidth]{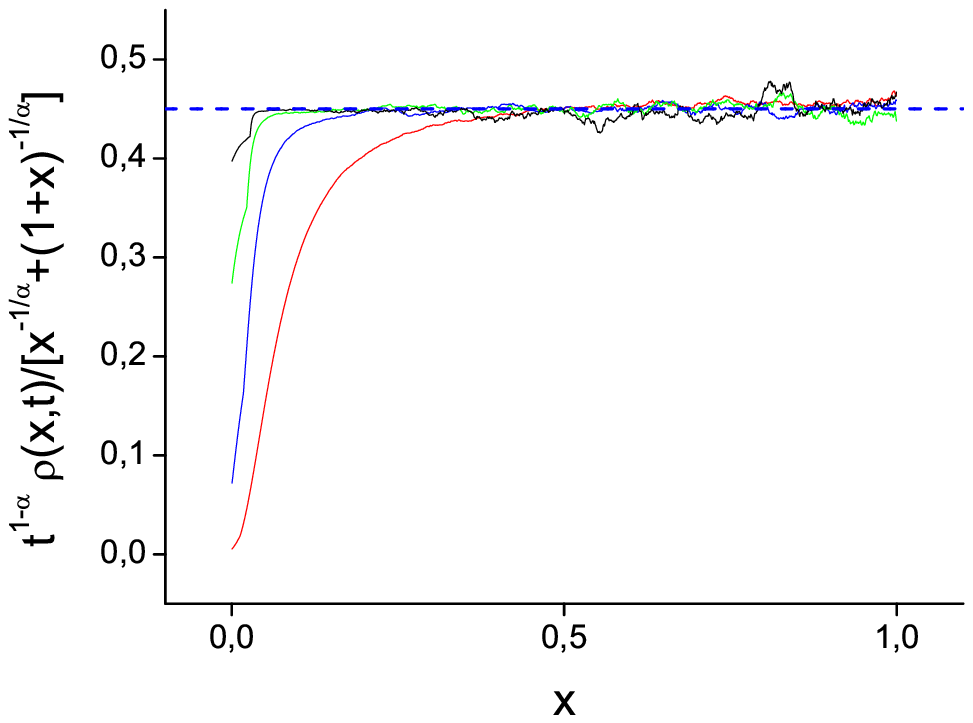}  
  \caption{Same as in figure \ref{FIG2NEW1}, $t^{1-\alpha} \rho(x,t)$ divided by $x^{-1/\alpha} + (1+x)^{-1/\alpha}$. Apparent convergence to a constant (dashed line given by equation (\ref{eqSVb}) without fitting) validates our numerical scheme. Noisy results for large times (upper curve counted from bottom to top on the left of the figure) and for large $x$ are due to statistical errors.}
  \label{FIG2NEW3}
\end{figure}

For the map (\ref{Thalereq}) Thaler has found an exact analytical expression for its infinite invariant density \cite{Thaler00}
\begin{equation}
\overline{\rho}(x_t) = B \left[ x^{-1/\alpha} + (1+x)^{-1/\alpha} \right].
\label{Thalerinf}
\end{equation}
Hence, unlike the PM map where we do not have an exact expression for the infinite density, for the map (\ref{Thalereq}) we can compare simulations with theory in the regime $0<x<1$. Since, as we mentioned for $x_t \rightarrow 0$ this map has the same behavior as the PM map, the constant $B$ is given by (\ref{eqSVb}). Note that the multiplicative constant $B$ is related to our working definition (\ref{eq3}) (see further discussion below). In figures \ref{FIG2NEW1} we see that $t^{1-\alpha} \rho(x,t)$ slowly converges towards the theoretical infinite density, besides the mentioned deviations close to $x \rightarrow 0$. As we increase measurement time the domain $x<x_c$, where deviations from asymptotic theory are observed is diminishing.  In figure \ref{FIG2NEW3} we plot $t^{1-\alpha} \rho(x,t)$ divided by $x^{-1/\alpha}+(1+x)^{-1/\alpha}$ showing that it converges to the constant $B$ as predicted in (\ref{Thalerinf}). This implies that our method of estimation of the infinite density works well, and hence we are confident it can be used also for maps where we do not have an exact expression for invariant density. We will use Thaler's analytical expression for the invariant density (\ref{Thalerinf}) to corroborate the generalization of the Pesin identity below. But first we briefly review the generalized Pesin identity.

\section{Generalized Pesin Identity}

Pesin's theorem, valid for $1<z<2$ where $\alpha =1$ asserts the equality 
\begin{equation}
\label{eq4}
\lambda_1 = h_{KS},
\end{equation}
where the Kolmogorov-Sinai entropy is given in terms of 
\begin{equation}
h_{KS}= \int_0 ^1 d x \ln|M'(x)| \; \rho_{\rm eq} (x).
\end{equation}
Here $\rho_{\rm eq}(x)$ is the normalizable invariant density. Pesin's identity provides a deep relation between chaotic and statistical quantities of the system. 

For $\alpha<1$ a different behavior is found. 
From (\ref{eq2}) and (\ref{eq3}) we suggested the generalization of the Pesin's identity in the form \cite{KB09,KB10}
\begin{equation} 
\left< \lambda_{\alpha} \right> = {1 \over \alpha} h_{\alpha},
\label{eq5}
\end{equation}
where the Krengel entropy appears 
\begin{equation}
h_{\alpha} = \int_0 ^1 d x \ln|M'(x)| \; \overline{\rho}(x).
\label{eq6}
\end{equation}
Note that R.~Zweim\"uller has shown the relation between the Krengel entropy and complexity $C_t$ \cite{Zwei}, so already at this point our work indirectly relates between the latter and separation of trajectories. Below we calculate the complexity using well-known compression algorithm of Lempel and Ziv and relate it to the Krengel entropy and $\left< \lambda_{\alpha} \right>$. Previously Lempel-Ziv complexity for weakly chaotic maps was studied in \cite{Arg02,SA06}.

Generalized Pesin identity equation (\ref{eq5}) follows from definitions of the generalized Lyapunov exponent (\ref{eq2}) and the infinite invariant density (\ref{eq3}) (which points out that with respect to the generalization of the Pesin identity these two definitions are related to each other). Using (\ref{eq2}) we have
\begin{equation}
\label{L_mean}
\left< \lambda_{\alpha} \right> = \int_0^1 \frac{\sum_{i=0}^{t-1} \ln \left| M'(x_i) \right|}{t^{\alpha}} \; \rho(x_0) \; dx_0,
\end{equation}
where the averaging is over initial conditions distributed according to a smooth initial density $\rho(x_0)$. Since we are interested in the long time limit we replace the summation with an integral and average over the density function 
\begin{equation}
\label{L_mean_1}
\left< \lambda_{\alpha} \right> \simeq \frac{1}{t^{\alpha}} \int_0^1 dx \int_{0}^{t} \ln \left| M'(x) \right|\; \rho(x,\tau) \; d \tau,
\end{equation}
where $\simeq$ underlines that this results is valid in the long time limit. Using (\ref{eq3}) we write the density as
\begin{equation}
\rho(x,t) = \lim_{t\rightarrow \infty} t^{\alpha-1} \overline{\rho}(x).
\end{equation}
Substituting this expression into (\ref{L_mean_1}) and performing integration over the time we arrive at 
\begin{equation}
\label{L_mean_c}
\left< \lambda_{\alpha} \right> = \frac{1}{\alpha} \int_0^1 dx \; \ln|M'(x)| \; \overline{\rho}(x),
\end{equation}
which is the generalization of the Pesin identity (\ref{eq5}), (\ref{eq6}) since the integral on the left is the Krengel's entropy. Notice that for $x \to 0$ the divergence of $\overline{\rho}(x)$ is canceled by $\ln|M'(x)|$. The $1/\alpha$ prefactor stems from the integration over time $t^{-\alpha} \int_{c}^{t} \tau^{\alpha-1} d\tau \rightarrow 1/\alpha$, where the constant $c$ regularizes the time integral (since we consider only long times). The constant $1/\alpha$ is a direct consequence of our definitions of generalized Lyapunov exponent and infinite invariant density and of course it can be absorbed in either definitions for aesthetics. We however stay with (\ref{eq5}), (\ref{eq6}) since the usual Lyapunov exponent is zero and the $1/\alpha$ serves as a reminder that we are dealing with a weakly chaotic system.

To clarify and avoid further confusion the $1/\alpha$ will appear in other averages. An important example is the complexity $C_t$ considered by Zweim\"uller \cite{Zwei}. Using the ratio ergodic theorem for complexity \cite{Zwei} we get complexity using our notations
\begin{equation}
\label{Ct}
C_t \rightarrow \frac{1}{\alpha} \int_0^1 dx \; \ln|M'(x)| \; \overline{\rho}(x).
\end{equation}
The $1/\alpha$ stems again from the summation (integration) over time, similar to what was done in (\ref{L_mean_1}). We emphasize as usual that this relation is valid for the definition (\ref{eq3}). Clearly we have \cite{KB10,ERRATUM}
\begin{equation}
\label{GP}
C_t = \left< \lambda_{\alpha} \right> = h_{\alpha}/\alpha.
\end{equation}

\begin{figure}[t!]
  \centering
  \includegraphics[width=.7\textwidth]{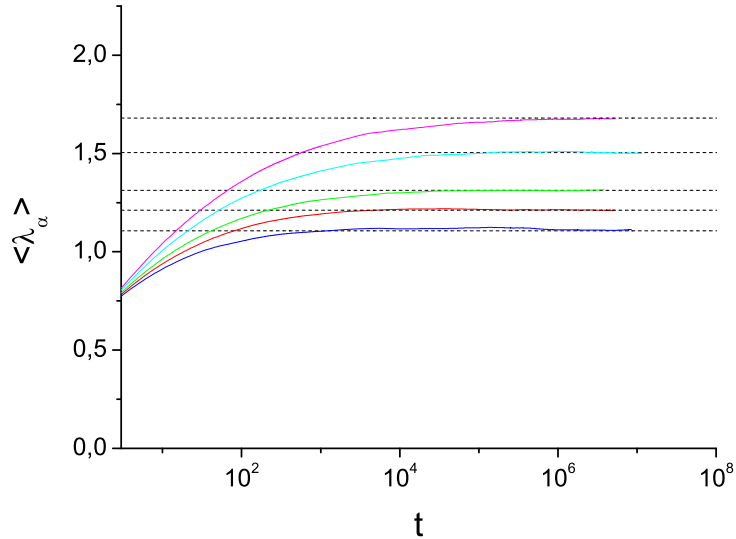}  
  \caption{Validation of the generalized Pesin identity (\ref{eq5}), (\ref{eq6}). Parameters of the Thaler map (\ref{Thalereq}) are $z=2.5, 2.55, 2.6, 2.7, 2.8$ (from bottom to top in the figure). The generalized Lyapunov exponent is calculated numerically according to (\ref{eq2}) starting from uniform ensemble. Dashed limiting lines correspond to the integral $\int_0^1 dx \; \ln|M'(x)| \; \overline{\rho}(x)/\alpha$ with exact analytical infinite invariant density (\ref{Thalerinf}) calculated using Mathematica program and constant $B$ given by (\ref{eqSVb}).} 
  \label{FIG2NEW4}
\end{figure}

\begin{figure}[t!]
  \centering
  \includegraphics[width=.7\textwidth]{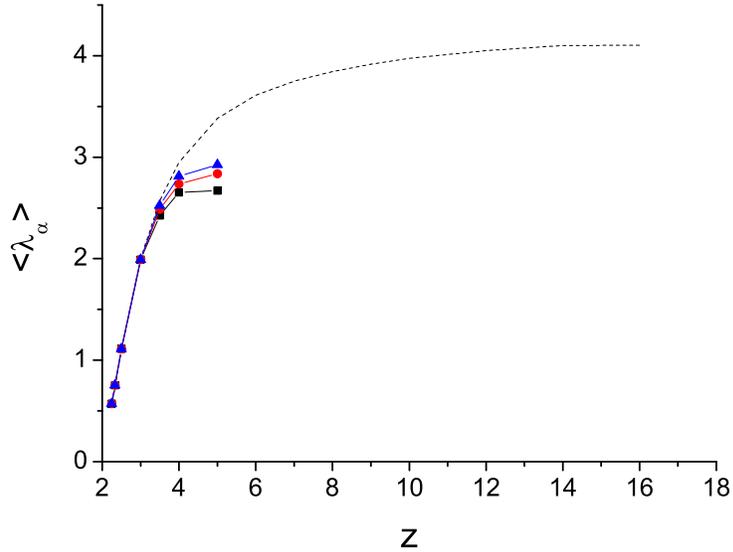}  
  \caption{The generalized Lyapunov exponent for the map (\ref{Thalereq}) is calculated numerically according to (\ref{eq2}) starting from a uniform ensemble of $10^5$ trajectories. Simulation time was $t=10^5$ (squares), $t=5 \; 10^5$ (circles) and $t=10^6$ (triangles). Notice slow convergence to theoretical curve. Dashed line corresponds to the integral $\int_0^1 dx \; \ln|M'(x)| \; \overline{\rho}(x)/\alpha$ with exact analytical infinite invariant density (\ref{Thalerinf}) calculated using Mathematica program and constant $B$ given by (\ref{eqSVb}).} 
  \label{FIG2NEW5}
\end{figure}

Now we use the exact analytical infinite invariant density (for the map $3$) to corroborate generalization of Pesin identity. We calculate the generalized Lyapunov exponent numerically according to (\ref{eq2}) starting from uniform ensemble. Using Mathematica we calculate the integral $\int_0^1 dx \; \ln|M'(x)| \; \overline{\rho}(x)/\alpha$ with the exact analytical infinite invariant density (\ref{Thalerinf}) with $B$ in (\ref{eqSVb}). Figure \ref{FIG2NEW4} fully corroborates our assertions with good accuracy. We note that convergence of numerics strongly depends on the parameter $z$  or $\alpha$. For large $z$ (corresponding to small $\alpha$) and for $z \rightarrow 2$ ($\alpha \rightarrow 1$) convergence of numerical results slows down. This is shown in figure \ref{FIG2NEW5}. However, for intermediate $z$ our generalized Pesin identity is fully supported by numerics. We note that here no stochastic approximation was employed whatsoever.

Clearly, our work provides the sought after elegant connection between entropy and separation of trajectories, at least for systems with an infinite invariant density. Operationally, however, what is our work about? It states that starting with a reasonable initial condition, e.g. particles uniformly distributed but not a delta function initial condition, the density of particles $\rho(x,t)$ evolves according to the transformation $M(x_t)$ and in the long time limit one can deduce from $\rho(x,t)$ the infinite invariant $\overline{\rho}(x)$ density using (\ref{eq3}). This can be done on a computer rather easily, or semi-analytically as we showed in our work \cite{KB09,KB10} (see also equation (\ref{rho1}) and (\ref{eqSVb})). Note that in this procedure we need not gather full information on the paths, only on their position after a long time $t$. On the other hand one can follow the trajectories of the particles and evaluate the sub-exponential separation using (\ref{eq2}). Thus, two numerical protocols are used one to find the density of particles and the other follows trajectories and measures the generalized Lyapunov exponent. With the infinite invariant density obtained in the first program, we can evaluate the Krengel entropy by performing an integral. This entropy is then used to evaluate the average separation. Of course our theory is testable in the sense that one can evaluate the infinite invariant density and with it predict the average separation, 
namely follow two different numerical protocols and checking the validity of our results (see below). 
We also found the fluctuations of the separation, our work being consistent with the Aaronson-Darling-Kac theorem.  Importantly, in \cite{KB10} we discussed the connection between the Krengel entropy and Lempel-Ziv
complexity thus showing the deep relation between
sub-exponential separation of trajectories and algorithmic complexity 
for weakly chaotic systems (see also discussion below). 
 
\subsection{ {\em Entities should not be multiplied unnecessarily} William 
of Occam} 

 One may claim that instead of using the definition (\ref{eq3})
for the infinite invariant density one could have used
another definition. One may suggest 
\begin{equation}
\overline{\rho}(x) = \lim_{t\rightarrow \infty} \frac{\rho(x,t)}{ C t^{\alpha-1}}
\end{equation}
with some $C>0$. 
In our work we choose $C=1$, which is consistent with the usual choice
of the normalized invariant density (the case $\alpha=1$)
\begin{equation}
\rho_{eq}(x) = \lim_{t \to \infty} {\rho(x,t) \over t^{\alpha -1} } |_{\alpha=1}.
\end{equation} 
The essence of the criticism on our work by SV \cite{SV} is that we could have chosen a different $C$.
However that would amount with only trivial multiplication of our
final results with a constant. Our results are valid as long
as one pays attention to our definitions, in particular
one should not ignore equation (\ref{eq3}). By not informing their reader that we use equation (\ref{eq3}), SV have distorted the context of our work.

Clearly, as we discussed above, our generalized Pesin identity depends also on the definition of the generalized Lyapunov exponent (\ref{eq2}). Similarly to the infinite invariant density it can be defined with some constant which we choose to be unity. However, as it follows from the derivation after (\ref{eq5}), (\ref{eq6}) choosing another constant in the definition of the generalized Lyapunov exponent would result in multiplication of our generalized Pesin identity by a trivial constant. 

The definition of these constants become important when considering averages. For example the generalized Lyapunov exponent converges to a constant 
\begin{equation}
\label{L2}
\left< \lambda_{\alpha} \right> = \frac{1}{\alpha} \int_0^1 dx \; \ln|M'(x)| \; \overline{\rho}(x).
\end{equation}
Because the left-hand side does not depend on the choice of an invariant density, namely it cannot depend on an arbitrary multiplicative constant, the infinite density $\overline{\rho}(x)$ must be determined precisely. Indeed exactly for this reason we must have $C=1$, and the working definition (\ref{eq3}) should not be ignored. One could of-course absorb the $1/\alpha$ in (\ref{L2}) in the definition of $\overline{\rho}(x)$, but as mentioned this is a matter of choice, which does not influence the predictive power of the theory.
 
\begin{figure}[t!]
  \centering
  \includegraphics[width=.7\textwidth]{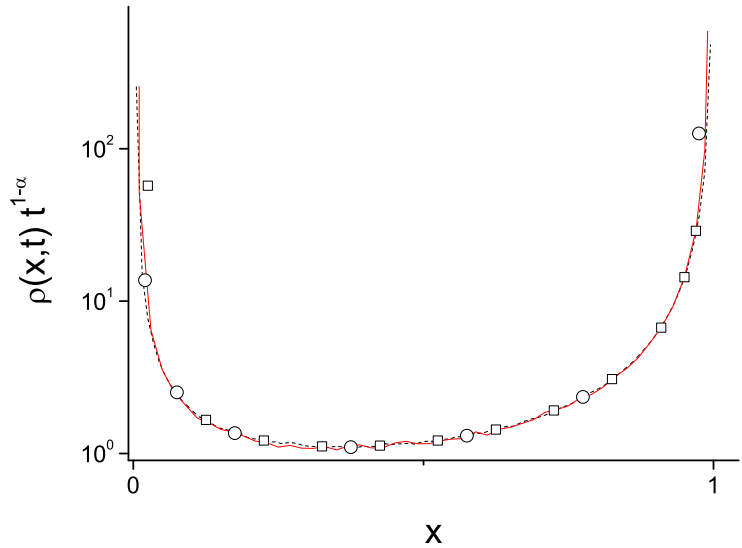}
  \caption{ $t^{1-\alpha} \rho(x,t)$ equation (\ref{eq3}) for the map (\ref{Man2}) with $c=0.3$ and $z=7/3$ ($\alpha=0.75$) calculated at $t=10^4$ (dashed line), $t=10^5$ (solid line) with uniform initial density $\rho(x,0)=1$ for $x \in (0,1)$. Similar results are
obtained for other initial conditions:  $\rho(x,0)=2$ in $x \in (0.5,1)$ (circles, $t=10^4$), and $\rho(x,0)=2 x$ in $x \in (0,1)$ (squares $t=10^5$).}
  \label{FIG1A}
\end{figure}

In their equation (16) SV claim to correct our result (\ref{eq5}). This boils down to other choices of $C$ and hence in our opinion is not a correction at all. In their work SV use an infinite invariant measure for the 
PM map $\omega(x)\sim B_{SV} x^{-1/\alpha}$, where $B_{SV}$ is an undefined constant 
(see their equation (2) and compare in our 
notations to $\overline{\rho}(x)$ equations (\ref{rho1}), (\ref{eqSVb})).
 Since they do not fix the constant $B_{SV}$ this measure is not 
the same as ours (freedom in choice of multiplicative constant $C$). Their Pesin-type identity reads 
\begin{equation}
\label{eqSV}
{h_\alpha \over B_{SV}} = a \left( {\alpha \over a } \right)^\alpha {\pi \alpha \over  \sin \pi \alpha} \langle \lambda_\alpha \rangle,
\end{equation}
where $h_\alpha$ is the Krengel entropy, with respect to invariant measure 
$\omega(x)$. 
Notice, that by fixing the constant $B_{SV}$ as in our equation (\ref{eqSVb}), SV relation (\ref{eqSV}) boils down to our generalized Pesin identity (\ref{eq5}). Also notice that SV presentation is specific to the PM map which has one unstable fixed point. Of course more generally we can have two or more unstable fixed points. The invariant density will reveal singularities next to $N$ unstable fixed points all with the same order of singularities (i.e. the same $z$) located on $b_i$ with $i=1, ..., N$ \cite{KB10,KB12,SA06}.
 For example $b_1=0$ and $b_2=1$ for the map (\ref{Man2}). 
 Then the infinite invariant density will be of the type
\begin{equation}
\label{eq15e}
\overline{\rho}(x) = \sum_i B_i \left| x - x_i \right|^{-1/\alpha} + o\left( \left| x - x_i \right|^{-1/\alpha} \right), \; \; x_i\rightarrow b_i.
\end{equation}
Moreover, for $c \ne 1/2$ the map (\ref{Man2}) is asymmetric with respect to $x=1/2$,  therefor the infinite invariant density is also asymmetric as shown in figure \ref{FIG1A} and $B_0 \ne B_1$. The identity suggested by SV does not describe this situation since it is restricted to a single unstable fixed point $N=1$ and hence is not general. In contrast, our Pesin-type identity is general and valid 
for maps with any number of unstable fixed points. So, we suggest 
to stick with our original identity (\ref{eq5}),
which is general and at the same time not distort its 
meaning by ignoring equation (\ref{eq3}). 

 In the second part of their work SV try to fix their claims. 
They write: {\em A closer inspection of our work (see in particular, their Eq.\ (10)) shows that they, when dealing with the continuous time stochastic 
linear model proposed in \cite{Grigo}, tacitly choose} 
\begin{equation}
 B = \left({ a \over \alpha} \right)^{\alpha-1} {\sin \pi \alpha \over \pi \alpha}. 
\end{equation}
 This statement is  misleading. In our work we explicitly give examples
of maps with two unstable fixed points, where this relation is
obviously wrong. In these maps one has two singularities in the infinite
density, so the infinite density is of the type 
\begin{equation}
\overline{\rho}(x) = \left\{
\begin{array}{ l  l}
B_1 x^{-1/\alpha} + o\left(x^{-1/\alpha}\right), & x \rightarrow 0, \\
B_2 (1 - x)^{-1/\alpha} + o\left((1 - x)^{-1/\alpha} \right), & x \rightarrow 1. 
\end{array}
\right.
\end{equation} 
Here we have two $B$'s, which, as we discussed above, for asymmetric maps are non identical. So, choosing a specific $B$ does not make any sense at all. We emphasize that our results are general and they are not sensitive to the choice of $B$, neither are they specific for one particular map. Rather we use equation (\ref{eq3}) which makes our results general while equation (\ref{eqSV}) is not.  

\begin{figure}[t!]
  \centering
  \includegraphics[width=.7\textwidth]{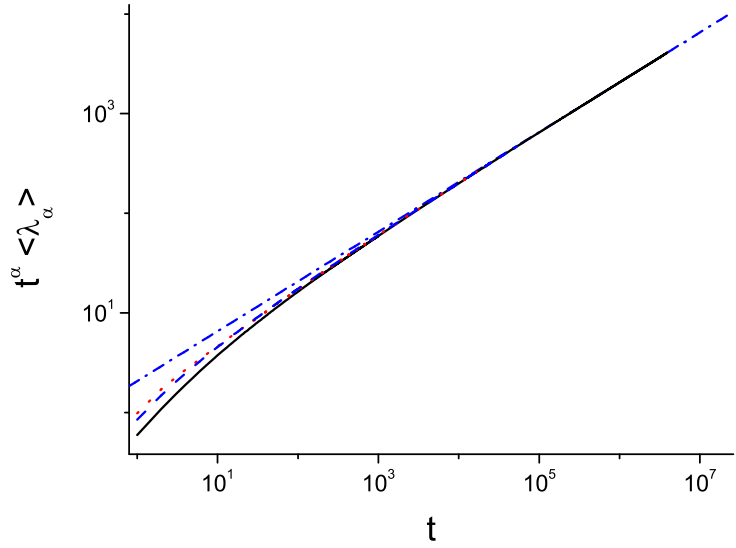}
  \caption{$t^{\alpha} \left< \lambda_{\alpha} \right> = \left< \sum_{k=0}^{t-1} \ln |M'(x_k)| \right>$ for the Pomeau-Manneville map
with $z=3$ ($\alpha=0.5$) calculated with different initial conditions: solid line $\rho(x,0)=1$ for $x \in (0,1)$, dashed line $\rho(x,0)=2$ for $x \in (0.5,1)$, dotted line $\rho(x,0)=2 x$ for $x \in (0,1)$. The dashed-dotted line is $(h_\alpha/\alpha) \; t^{\alpha}$ where the entropy  $h_{\alpha}/\alpha\simeq 2.06$ 
is calculated  using numerical integration of (\ref{eq6}).}
  \label{FIG2}
\end{figure}
\begin{figure}[t!]
  \centering
  \includegraphics[width=.7\textwidth]{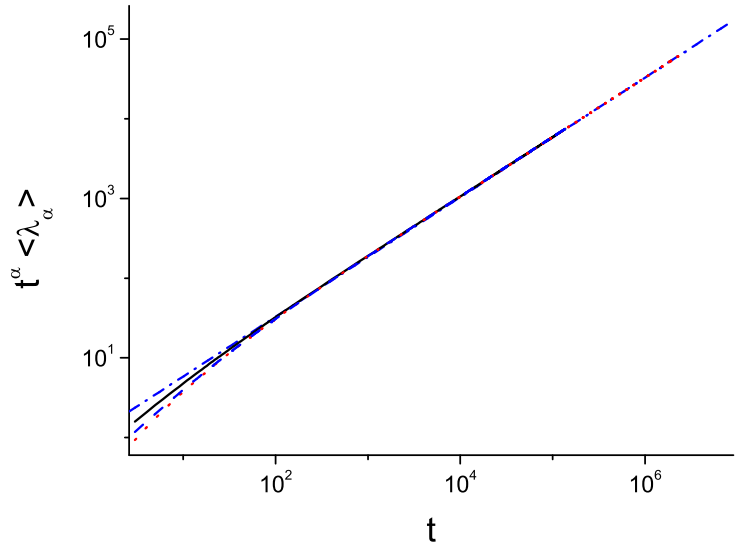}
  \caption{$t^{\alpha} \left< \lambda_{\alpha} \right> = \left< \sum_{k=0}^{t-1} \ln |M'(x_k)| \right>$ for the map (\ref{Man2}) with $c=0.3$ and $z=7/3$ ($\alpha=0.75$) calculated with different initial conditions: solid line $\rho(x,0)=1$ for $x \in (0,1)$, dashed line $\rho(x,0)=2$ for $x \in (0.5,1)$, dotted line $\rho(x,0)=2 x$ for $x \in (0,1)$. The dashed-dotted line is $(h_\alpha/\alpha) \; t^{\alpha}$ where  $h_{\alpha}/\alpha\simeq 1.038$ is found using (\ref{eq6}).}
  \label{FIG4}
\end{figure}

\begin{figure}[t!]
  \centering
  \includegraphics[width=.7\textwidth]{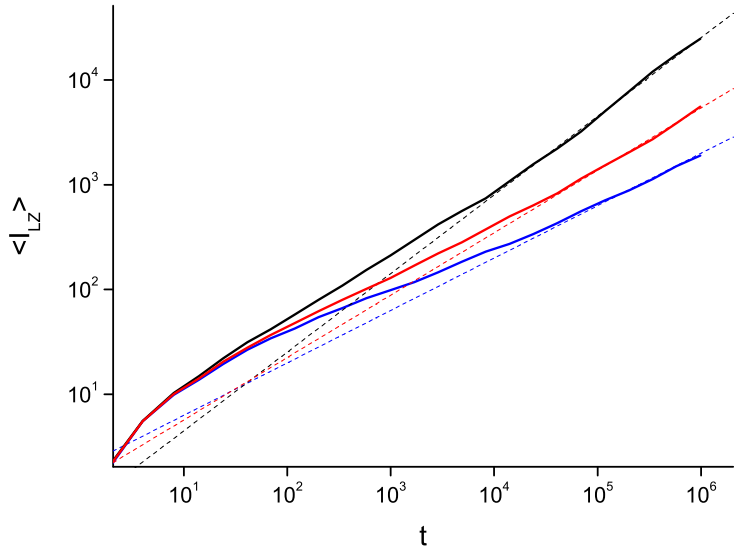}
  \caption{Information content of trajectories $ \left< I_{LZ} \right> = c(t) (\log_2 c(t) + 1) $ for the map (\ref{Man2}) (here $c(t)$ is the number of words in a trajectory, see \cite{KB10} for more details) calculated by the Lempel-Ziv algorithm for $z=3$, $z=2.68$ and $z=2.33$ (from bottom to top). Dashed lines correspond to $\left< I_{LZ} \right> = h_{\alpha} t^{\alpha}/\alpha$ with $h_{\alpha}$ found using (\ref{eq6}). Each curve is averaged over $100$ initial conditions.}
  \label{FIGLZ}
\end{figure}

Finally, in our work we corroborated Gaspard and Wang result \cite{GW88} that $\lambda_{\alpha}$ is proportional to the number of injections in vicinity of unstable fixed points $N_t$ (see more details in \cite{KB09,KB10}). We further showed that $N_t$ and hence $\lambda_{\alpha}$ are random variables with a Mittag-Leffler distribution in accordance with the Aaronson-Darling-Kac theorem. On this issue SV wrote: \textit{It is interesting to notice that $N_t$ is also considered as a Mittag-Leffler random variable in \cite{KB09,KB10} by using renewal theory in a different manner, but its relation to $C_t$ is not stated.} The second part of this sentence is problematic.  In the abstract of Ref.\ \cite{KB10} we wrote: \textit{We show that $\alpha \left< \lambda_{\alpha} \right>$ is equal to Krengel entropy and to the complexity calculated by the Lempel-Ziv compression algorithm}. This relation is given by \cite{KB10,ERRATUM}
\begin{equation}
C_{LZ} = \left< \lambda_{\alpha} \right> = h_{\alpha}/\alpha.
\end{equation} 
In figure \ref{FIGLZ} we repeat our numerical calculation of the average information content $\left< I_{LZ} \right>$ by the Lempel-Ziv compression algorithm for different values of $\alpha$ and for longer time (see \cite{KB10} for more details). We then calculate the Lempel-Ziv complexity as $C_{LZ} = \left< I_{LZ} \right> /t^{\alpha}$. Results of figure \ref{FIGLZ} are fully consistent with those in \cite{KB10}. In this proposition we suggest that $C_{LZ}$ is an estimator  of the complexity $C_t$. In equation (20) an exact relation between complexity and $\left<\lambda_\alpha\right>$ is given. Since $C_t$ is not directly computable, replacing $C_t$ in equation (20) with the estimator $C_{LZ}$ yields equation \ (28) which is not rigorous, and so far supported by numerical evidence only. The proposition is motivated by the fact that in the
non-zero entropy limit $\alpha>1$ the Lempel-Ziv complexity is a good estimator 
of $C_t$ \cite{TCbook}. Clearly more work in this direction is needed.

\section{Infinite invariant density obtained from different
initial states}

 To go beyond general relations and for the sake of specific
predictions  we need estimates for the infinite
invariant density. That goal is in principle rather simple.
 We start the evolution with initial conditions whose density does not contain a delta function, e.g. uniform initial conditions
and after long measurement time estimate $\overline{\rho}(x)$ 
using (\ref{eq3}). Numerically we now demonstrate that the infinite invariant density defined by (\ref{eq3}) does not depend on the choice of initial conditions. Namely we assume a normalizable 
initial density, not containing singularities (like delta functions). 

Three initial conditions are considered: $\rho(x,0)=1$ for $x \in (0,1)$, $\rho(x,0)=2$ for $x \in (0.5,1)$, and $\rho(x,0)=2 x$ for $x \in (0,1)$).
For these choices we   evolve $\rho(x,t)$  and then
 using Eq.\ (\ref{eq3}) we estimate $\overline{\rho}(x)$.
Supported by simulations we see that when $t \to \infty$ we get the same result 
for $\overline{\rho}(x)$ 
independent of the initial state. We have checked this for two maps (\ref{Maneq})  and  (\ref{Man2}) which have one and two unstable fixed points respectively. 
The results are shown in figures \ref{FIG1}, \ref{FIG1A}.
 We see that $\overline{\rho}(x)$ is independent of the initial
state.  This implies that one can attain an estimate for the infinite
 invariant
density rather easily, though more rigorous work is needed to give
estimates on the convergence rate. Our work also shows that  
the generalized Lyapunov exponent, can be estimated starting from 
different initial conditions. Of course at short times
the estimates will vary from one initial condition to another,
however as shown in figure \ref{FIG2} and \ref{FIG4} different initial states
give the same estimate for $\langle \lambda_\alpha \rangle$ in perfect agreement
with the generalized Pesin identity.

\section*{Summary}

In the mathematical literature the infinite invariant density is 
defined up to an arbitrary multiplicative constant.
We followed William of Occam economical philosophy  and 
we fixed the constant to unity. 
More practically, to test predictions of a theory we need 
estimates for the infinite invariant
density, which we obtain from a theory or numerics.
It is therefore useful to define the infinite density precisely as we did in equation (\ref{eq3}), and not leave it defined up to an arbitrary multiplicative constant. This operational definition is useful, since as we demonstrated it can be used to estimate the infinite density. With the infinite density we may calculate averaged observables and here we focused on $\left< \lambda_{\alpha} \right>$ which is a measure of sub-exponential separation. Here we used exact expression for infinite density to obtain $\left< \lambda_{\alpha} \right>$ which perfectly match simulations. Unfortunately exact expression for infinite invariant density are scarce, and hence we believe our numerical approach is useful. 
We showed that the criticism posed recently on our generalization 
of Pesin's identity for weakly chaotic systems is unjustified. 
We propose to stay with our identity because of its testability 
and broad validity, beyond the single unstable fixed point case.    

\section*{Acknowledgments}

This work is supported by the Israel Science Foundation.


\end{document}